\documentclass[fleqn,usenatbib]{mnras}

\usepackage{newtxtext,newtxmath}

\usepackage[T1]{fontenc}

\DeclareRobustCommand{\VAN}[3]{#2}
\let\VANthebibliography\thebibliography
\def\thebibliography{\DeclareRobustCommand{\VAN}[3]{##3}\VANthebibliography}

\usepackage{graphicx}	
\usepackage{amsmath}	
\usepackage{amssymb}	

\newcommand{\hii}    {H\,{\sc{ii}}}
\newcommand{\cii}    {C\,{\sc{ii}}}
\newcommand{\kms}    {\,km\,s$^{-1}$}
\newcommand{\co}     {CO(6--5)}
\newcommand{\tco}     {$^{13}$CO(6--5)}
\newcommand{\innersize}     {0.9~pc}

\title[The shocked molecular layer in RCW\,120]{The shocked molecular layer in RCW\,120}

\author[M. S. Kirsanova et al.]{
M. S. Kirsanova,$^{1}$\thanks{E-mail: kirsanova@inasan.ru}
Ya. N. Pavlyuchenkov,$^{1}$
A. O. H. Olofsson,$^{2}$
D.~A. Semenov,$^{3,4}$
A.~F. Punanova$^{5}$
\\
$^{1}$Institute of Astronomy, Russian Academy of Sciences, 119017, 48 Pyatnitskaya Str., Moscow, Russia\\
$^{2}$Department of Space, Earth and Environment, Chalmers University of Technology, Onsala Space Observatory, SE-43992 Onsala, Sweden\\
$^{3}$Max-Planck-Institut f{\"u}r Astronomie, K{\"o}nigstuhl 17, 69117 Heidelberg, Germany\\
$^{4}$Department of Chemistry, Ludwig Maximilian University, Butenandtstra{\ss}e 5--13, 81377 Munich, Germany\\
$^{5}$Ural Federal University, 620002, 19 Mira street, Yekaterinburg, Russia
}

\date{Accepted XXX. Received YYY; in original form ZZZ.}

\pubyear{2015}

\begin{document}
\label{firstpage}
\pagerange{\pageref{firstpage}--\pageref{lastpage}}
\maketitle

\begin{abstract}
Expansion of wind-blown bubbles or \hii{} regions lead to formation of shocks in the interstellar medium, which compress surrounding gas into dense layers. We made spatially and velocity-resolved observations of the RCW~120 PDR and nearby molecular gas with \co{} and \tco{} lines and distinguished a bright CO-emitting layer, which we related with the dense shocked molecular gas moving away from the ionizing star due to expansion of \hii{} region. Simulating gas density and temperature, as well as brightness of several CO and C+ emission lines from the PDR, we found reasonable agreement with the observed values. Analysing gas kinematics, we revealed the large-scale shocked PDR and also several dense environments of embedded protostars and outflows. We observe the shocked layer as the most regular structure in the \co{} map and in the velocity space, when the gas around YSOs is dispersed by the outflows.
\end{abstract}




\section{Introduction}\label{sec:intro}

Massive stars shape the Galaxy via radiative and mechanical feedback and appear as lighthouses of star formation. The feedback is primarily visible through optical emission of ionized (\hii) regions in the plane of the Galaxy \citep[e.~g. recent H-alpha surveys by][]{SHASSA, IPHAS}. Theory predicts that shock waves, related to expansion of the \hii{} regions, compress the neutral gas and dust and collect it into moving dense molecular layers, \citep[e.g.][]{Spitzer_1978}. One of the most studied dense shocked layers on the border of an \hii{} region is the Orion Bar photo-dissociation region (PDR, the region between ionization front and attenuated cold molecular gas), which has already been observed with high angular resolution in optical, infrared and millimetre wavelengths \citep[e.~g.][]{McLeod_2016, Goicoechea_2016, Berne_2022}. A steady-state structure of the Orion Bar PDR where dissociation fronts of different molecules are separated from each other by several values of $A_{\rm V}$ was developed by \citet{Tielens_1985,Tielens_1993}. Only after the appearance of the first ALMA data towards the PDR, dynamical effects have been revealed by \citet{Goicoechea_2016} and confirmed numerically by \citet{Kirsanova_2019b}. Since the Orion Bar PDR is a unique object due to its proximity, many other PDRs are all less studied. Moreover, retirement of the SOFIA telescope, which brought a bulk of important observational data towards PDRs, leads to difficulties in getting more data about PDRs. Therefore, our knowledge about PDRs are highly limited by the strong selection effect and every new study of their properties is important.

\begin{figure}
\includegraphics[width=\columnwidth]{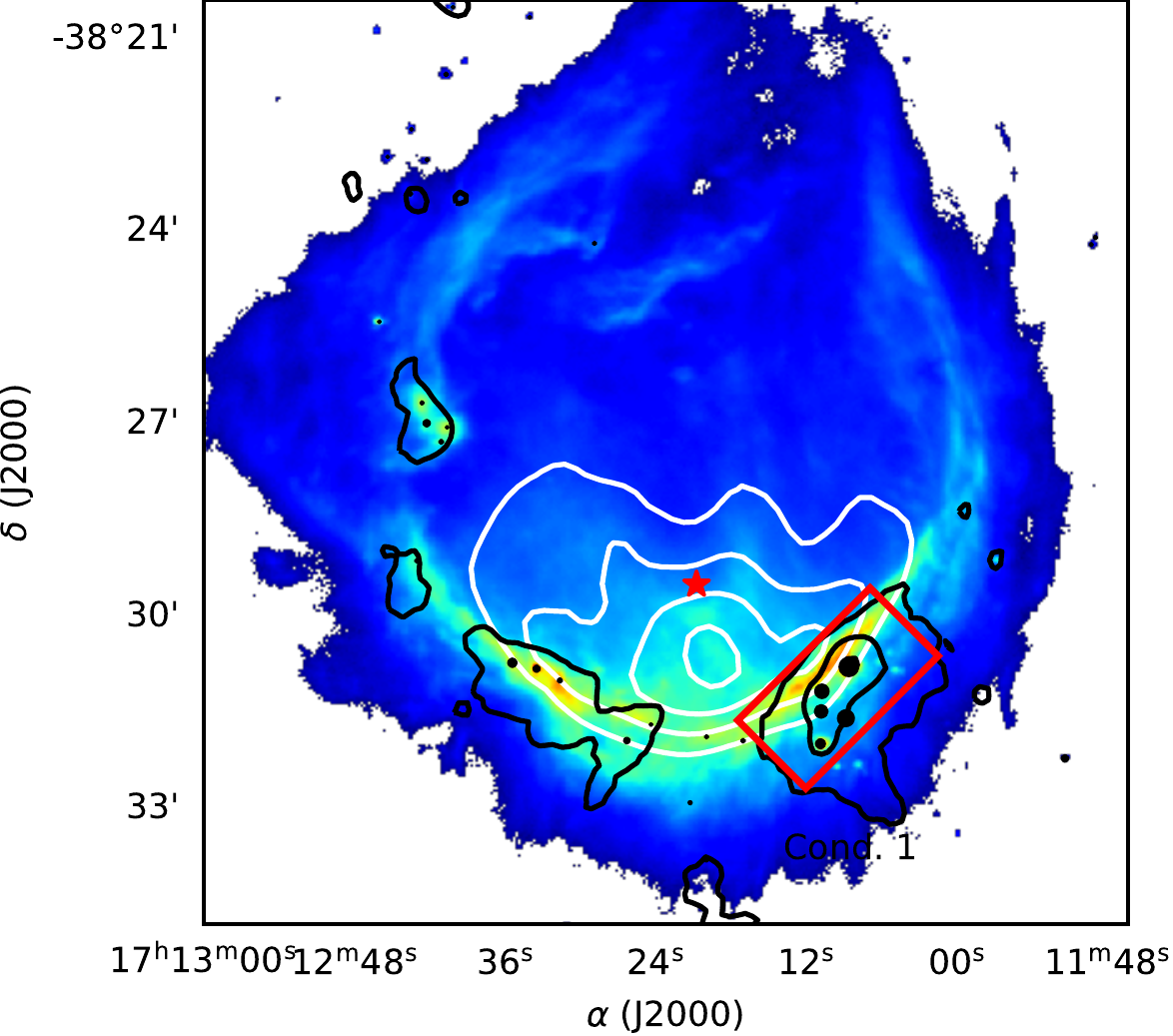}
\caption{Image of RCW\,120 at 70\micron{}. Color scale shows emission of neutral material around the ionized gas. The position of the ionizing star of RCW~120 is marked as a red star. The 843~MHz radio continuum emission corresponding to free-free emission of the ionized gas is shown with white contours linearly spaced from from 0.1 to 0.4 Jy/beam. The 870\,$\rm \mu$m contours for 0.4, 2.0 and 10.0 Jy/beam are shown in black. Black circles show the locations of compact sources described by \citet{Figueira_2017} (their Table~5). The sizes of the circles depend linearly on the source masses \citep[$M_{\rm env}$ from][]{Figueira_2017} here and on the following figures. The area, observed in the \co{} line, is shown by the red rectangle.}
\label{fig:obsregion}
\end{figure}

The southern \hii{} region RCW~120 is powered by an O6-8V/III ionizing star \citep{Martins_2010} and has a radius of 250\arcsec. Simple ring-like geometry of RCW~120, which resembles a projection of a 3D spherical shell, makes it attractive to study. However, the position of the ionizing star does not corresponds to the center of the neutral envelope of the \hii{} region, see Fig.~\ref{fig:obsregion}. Moreover, the maximum of the free-free continuum emission is shifted to the south up to 1.5\arcmin{} in the plane of the sky. Therefore, the simple geometry of RCW~120 might be illusory. RCW\,120 is one of the nearest \hii\ regions to the Sun, 1.34~kpc distant~\citep[][]{Russeil_2003}\footnote{We prefer not using 1.7~kpc from \citet{2019ApJ...870...32K} because GAIA gives the best distances for unobscured stars but the extinction $A_{\rm V} = 4.4$ towards RCW~120, see \citet{Zavagno_2007}. }. Physical radius of the \hii{} region is therefore about 1.6~pc but the mean projected distance between the ionizing star and the southern neutral border of the \hii{} region is about 1~pc. The cold neutral material around RCW\,120 consists of several condensations arranged to the shell-like structure, the condensations contain embedded young stellar objects (YSOs) \citep{Zavagno_2007,Deharveng_2009}. The most massive is a Class~0 YSO with $M_{\rm star}=8-10 M_{\odot}$ (source\,2 in ~\citet{Figueira_2017}, S2 below). Recently, \citet{Luisi_2021} found an expanding continuous shell around RCW\,120 due to a stellar wind impact, observing the \cii{} emission at 158~\micron. 

The neutral environment of RCW~120 has different geometry compared with the view at the \cii{} line. The PDR appeared in flattened face-on molecular cloud partly embedded into a low-density envelope \citep[see][]{Anderson_2015, Kirsanova_2019, Kabanovic_2022}. Therefore, RCW~120 is not a simple 3D sphere or 2D ring, but different parts of the object can be modelled by a 1D or 2D numerical model or by their combination. Modelling of the region as a whole can probably be done in some approximation such as a smoothed-particle hydrodynamics approach \citep[see][]{2015MNRAS.452.2794W}.

Since RCW~120 is expanding, one can expect to find the dense shocked layer in its PDR. While the mentioned above authors studied RCW~120 with spectral lines of CO and its isotopologues, none of these studies had sufficient spatial resolution to resolve the shocked layer. \citet{Kirsanova_2019} calculated the physical structure of the RCW~120 PDR and found a width of the dense moving layer of 0.1~pc. Using single-dish ground-based telescopes, this layer can be resolved only with the APEX telescope and \co{} lines at a frequency $\approx 660$~GHz \citep[Fig.~1 in][]{2018abmc.conf..284K} or by ALMA. In this study, we present the resolved observations of the dense shocked layer in RCW~120.
 
\section{Observational data and numerical simulations}\label{sec:obs}

We performed new observations of the RCW~120 PDR and combined their results with simulations of the region in order to distinguish between large-scale phenomena, related to the kinematics of the \hii{} region, and local star-formation processes, related to star-forming activity, on molecular emission maps.

\subsection{\co{} and \tco{} observations}

The observations were carried out using the Atacama Pathfinder EXperiment telescope (APEX) in Chile on 15 and 17/18 of October 2019 and 10 July 2021, as projects O-0103.F-9301A-2020 and O-0107.F-9318A-2021 (PI: Kirsanova M.~S.) within the Swedish operated share. The receiver used was the Band~9 part of the SEPIA bundle \citep{Baryshev_2015,Belitsky_2018}, also called SEPIA~660. In case of $^{12}$CO (2019), it was tuned to 691~GHz (the line frequency is 691473.076~MHz) covering the sideband ranges 685-693~\&~701-709~GHz, and for $^{13}$CO (2021) it was tuned to 661067.276~MHz covering 655-663 \& 671-679~GHz. The spectral resolution in the FFT spectrometer used was about 61~kHz (26~m~s$^{-1}$) with 65536 channels per every 4~GHz. The data were calibrated to antenna temperature in real-time using the standard {\it apexOnlineCalibrator} package, but we later additionally applied factors of $\eta_{\rm mb} = 0.40$ and~0.53 for \co{} and \tco, respectively\footnote{https://www.apex-telescope.org/telescope/efficiency/index.php}, to arrive at the main beam temperature scale. The spatial resolution of the data was 9\arcsec, corresponded to 0.058~pc at the distance of RCW\,120.

We observed a region with different physical conditions: from the irradiated side of molecular clump on the border of the ionization front (photo dissociation region, RCW\,120~PDR below) to the outskirts of the clump, see Fig.~\ref{fig:obsregion}. The region contains several massive YSOs. For $^{12}$CO, the region was mapped using the OTF mapping mode to cover 175\arcsec{} by 90\arcsec{} rotated 45$^\circ$\ in the equatorial system (the long side extending from southeast to northwest, see Fig.~\ref{fig:obsregion}). The rows/columns in the map were alternatingly observed along the x- and y-directions with a data dump time of 1.0 seconds and a step of 3.1\arcsec. The off-position was chosen at the direction $\alpha=17\rm ^h$12$\rm ^m$08.000$\rm ^s$, $\delta=-38^\circ$36\arcmin03.00\arcsec (J2000). The weather was rather poor with $0.9<{\rm PWV}<1.6$~mm (corresponding to $T_{\rm sys} \approx 1000-4000$~K) for the $^{12}$CO map given this atmospheric window but relatively stable and calibrations were inserted more often to ensure consistent intensity calibration.

The $^{13}$CO observations were carried out as an 90\arcsec{} OTF strip cutting through the peak integrated intensity position of the $^{12}$CO map at a right angle to the long axis of the map. The data dump time and step size were the same as for the $^{12}$CO observations. Total observational time was 1.7~h. The weather conditions were better for the $^{13}$CO observations with $0.6<{\rm PWV}<0.7$~mm and typical $T_{\rm sys} = 730$~K.

The gridding of the data and the baseline correction were performed using the CLASS package from the GILDAS\footnote{http://www.iram.fr/IRAMFR/GILDAS} software. Further analysis was done with Astropy \citep{astropy:2013, astropy:2018}, and APLpy \citep[][]{aplpy2012, aplpy2019} was used for representation. Final fits-cubes were proceeded to the common grid with the step size of 3.1\arcsec{} in x- and y-directions. Typical noise level ($1\sigma$) of the regridded data is 1.5-1.7~K.

\subsection{Archival data}

We used the multi-wavelength data from various archives. In the section, we briefly describe them and provide needed references. The {\it Herschel} HIGAL data at 70~\micron{} \citep{2010PASP..122..314M} and the SUMSS 843~MHz radio continuum emission \citep[][]{Bock_1999} were used to outline general geometry of the neutral environment RCW~120 and the ionized gas, respectively. The ATLASGAL 870~\micron{} emission was used to locate dense and cold molecular clumps \citep{2009A&A...504..415S}. The [\cii] line emission at 158~\micron{} obtained as a part of the FEEDBACK Legacy Program \citep[][]{2020PASP..132j4301S} were downloaded from the {\it SOFIA} archive and used to study the irradiated side of the RCW~120 PDR together with the {\it Spitzer} data at 8~\micron{} \citep[][]{Zavagno_2007}.  We also used the $^{12}$CO and $^{13}$CO (3--2) data \citep{Kabanovic_2022} to study dependence of our conclusions on spatial resolution.

\subsection{Numerical simulations}

We simulated the physical and chemical structure of RCW~120 using the MARION model according to the approach by \citet{Kirsanova_2019}. In the present study, all the model parameters were taken the same as those authors did. We took the model results at the moment $5.8\times10^5$~yrs from the beginning of the \hii{} region expansion when the \hii{} region has the radius about 1~pc. We note that this model time should not be considered as an age of RCW~120, because geometry of the region is not spherical and density distribution is not uniform. However, our 1D model can be used to study the shocked region on a scale of a particular molecular clump because we are not pretending to simulate the line profiles, related to geometry, but interested only in the line brightness, related to the energy of the region. Using the simulated physical and chemical structure, we applied RADEX software~\citep[][]{vanderTak_2007} in order to model brightness of the various lines of CO, $^{13}$CO and \cii. Coefficients for collisions with H$_2$ were used for all considered emission lines because this molecule is the most abundant collision partner in our model. Using APEX beamwidth at 1.3, 0.9 and 0.4~mm (27\arcsec, 18\arcsec and 9\arcsec) as well as {\it SOFIA} beamwidth at 158~micron (15\arcsec), we simulate the brightness of molecular lines from RCW~120 for those telescopes.

\section{Results}\label{sec:res}

\subsection{Observed \co{} and \tco{} line emission}\label{sec:geom}

\begin{figure*}
\includegraphics[width=0.99\columnwidth]{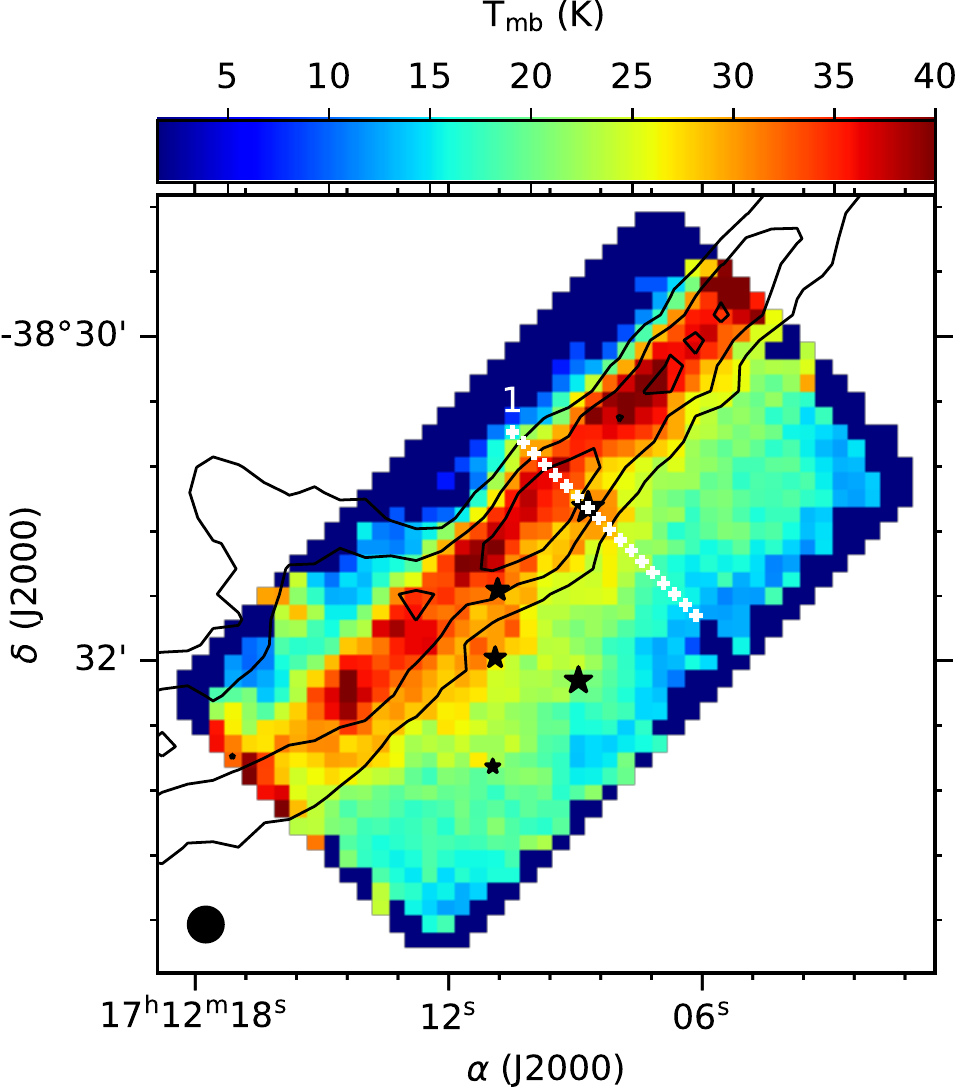}
\includegraphics[width=0.99\columnwidth]{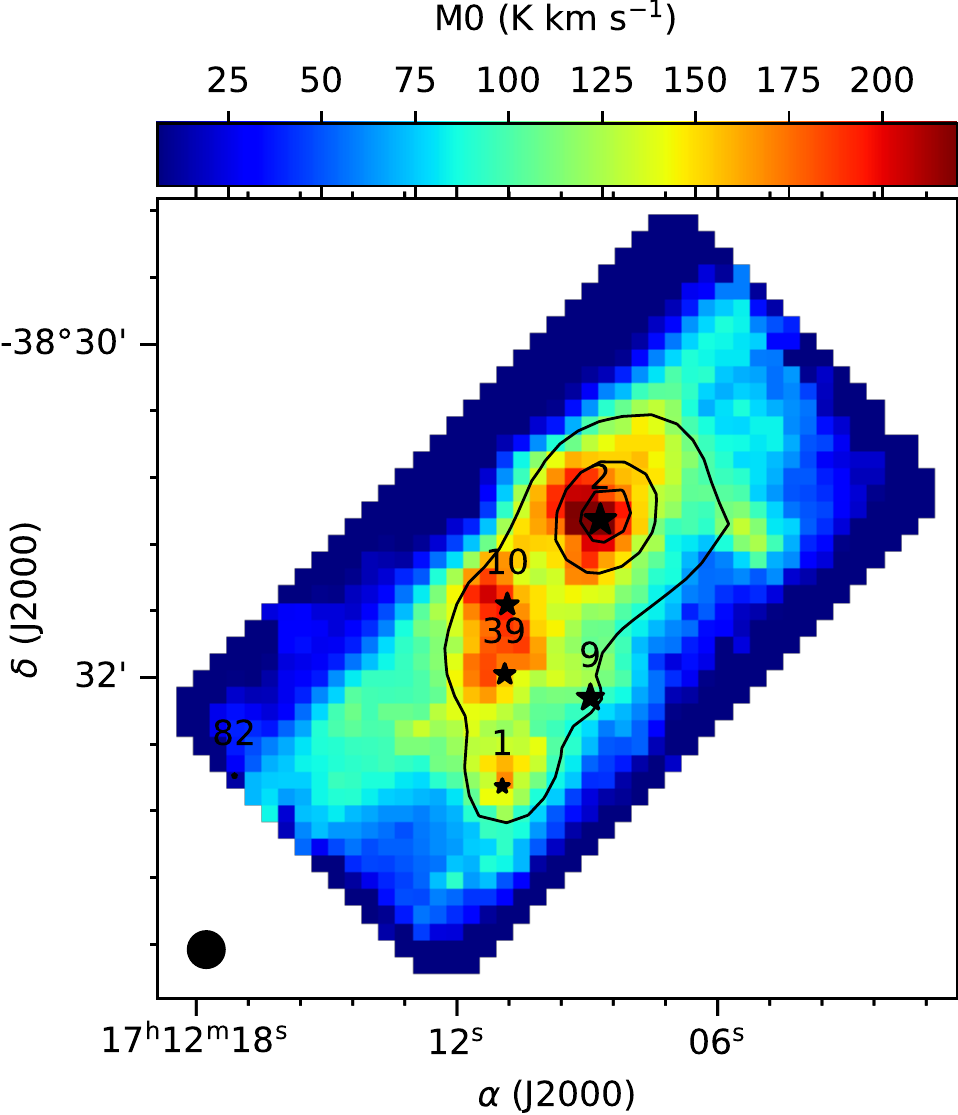}
\caption{Left: map of the \co{} peak intensity in the RCW~120 PDR shown by colour scale. Contours show brightness of the [\cii] line emission at 20, 30 and 40~K. White crosses show positions of the \tco{} spectra shown in Fig.~\ref{fig:individual_spectra}, where position~1 corresponds to illuminated side of the PDR. Right: map of the \co{} integrated intensity shown by colour scale. The 870\,$\rm \mu$m contours for 0.4, 2.0 and 10.0 Jy/beam are shown by contours. The APEX beam at 691~GHz is shown by black circle. The black stars show the locations of YSOs from \citet{Figueira_2017} in both plots and in following figures.}
\label{fig:obsemission}
\end{figure*}

Spatial distributions of the \co{} peak and the integrated intensity are different as they are defined by large-scale and local phenomena, respectively. Namely, the peak intensity of the \co{} emission in Fig.~\ref{fig:obsemission} has a layer-like spatial distribution coinciding with the [\cii{}] emission in the RCW\,120 PDR. The peak \co{} intensity drops sharply on the irradiated side of the PDR but decreases smoothly on the attenuated side. The layer of the bright \co{} emission appears parallel to the ionization front (compare Fig.~\ref{fig:obsregion} and~\ref{fig:obsemission}) and does not become warped around the border of the dense and cold material, traced by ATLASGAL 870\,$\mu$m contours (compare left and right panels in Fig.~\ref{fig:obsemission}). Therefore, the bright CO-layer represents a shell of irradiated gas related to large-scale gas distribution created by RCW\,120.

\begin{figure}
\includegraphics[width=0.9\columnwidth]{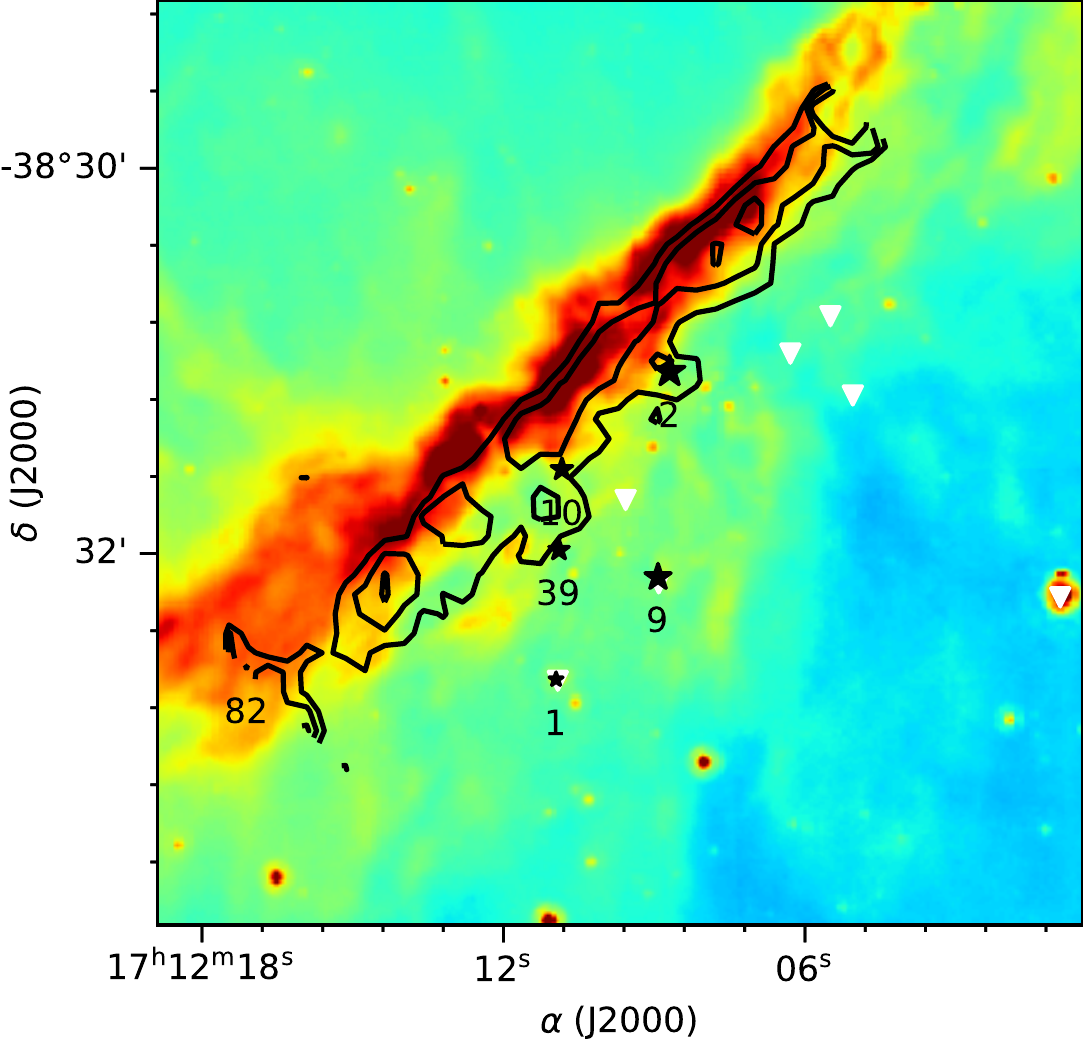}
\caption{Image at 8\micron{} and the map of the \co{} peak intensity are shown by colour and contours, respectively. The contours are shown for 30, 35 and 40~K\kms. 
White triangles show YSOs from \citet{Deharveng_2009}. }
\label{fig:spitzer}
\end{figure}

The shortest projected distance from the ionizing star (shown in Fig.~\ref{fig:obsregion}) to the bright CO-layer is 146.5\arcsec, corresponding to \innersize. Physical width of the layer, in the direction perpendicular to the ionization front, is $\approx 0.10-0.15$~pc as we deduce from spatial distribution of the peak intensity in Fig.~\ref{fig:obsemission}. Note that the width of the bright CO-layer remains almost the same over the entire region we observed in contrast with the [\cii{}] and 8\micron{} emission, shown in Fig.~\ref{fig:obsemission} and \ref{fig:spitzer}, whose widths become broader up to a  factor of 1.5-2 in the south-eastern part of the observed region. Two of five massive YSOs, namely S2 and S10, are projected on the bright CO-layer, the other three massive YSOs are located outside the layer.  Several point sources, found by \cite{Deharveng_2009}, (see Fig.~\ref{fig:spitzer}), are located outside the CO-layer. Width of the whole molecular clump perpendicular to the ionization front is $\approx 0.4$~pc.

On the other hand, the map of the integrated intensity reveals also rather local phenomena, related to regions of star-forming activity. Several peaks of the integrated \co{} intensity appear towards YSOs but not towards the PDR's edge. Therefore, broadening of the \co{} lines towards the YSOs is responsible for the appearance of the integrated intensity peaks. The highest integrated intensity of \co{} coincides with the peak of the ATLASGAL 870~$\mu$m emission and the most massive YSO\,S2. There are three minor peaks of the \co{} emission towards YSOs\,S1, S10 and S39. There is one more massive YSO\,S9 in the map, but it does not demonstrate brightening of the \co{} emission. 

\begin{figure*}
    \centering
    \includegraphics[width=2.\columnwidth]{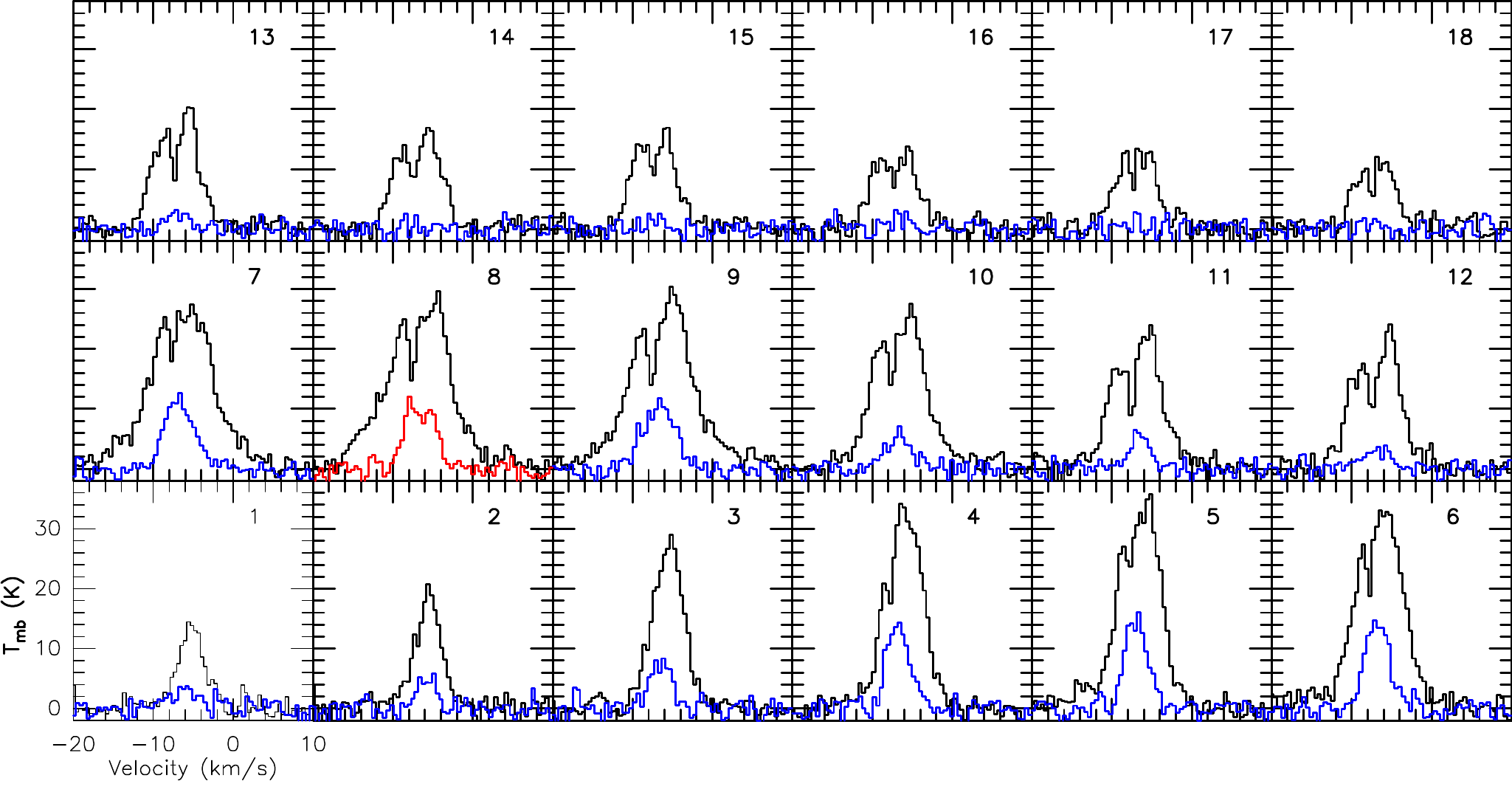}
    \caption{\co{} (black) and \tco{} (blue) spectra along the strip shown in Fig.~\ref{fig:obsemission}. The spectra are shown with the step of 3\arcsec{} from the north-east (position 1 at the offset $\Delta\alpha, \Delta\delta =$~21\arcsec,21\arcsec) to south-west (position 18 at the offset $\Delta\alpha, \Delta\delta =$~--30\arcsec,--30\arcsec). Towards S2, the \tco{} line is shown by red colour (position 8).}
    \label{fig:individual_spectra}
\end{figure*}

The regions affected by the large-scale (related with the expanding \hii{} region) and local (related with the star-forming activity towards the YSOs) phenomena are clearly distinguishable in Fig.~\ref{fig:individual_spectra}, where we show individual \co{} and \tco{} spectra. The \co{} lines are single-peaked in positions~1-4 on the irradiated side of the PDR. Deeper to the molecular clump, the line profiles become double-peaked, less symmetric and demonstrate pronounced red and blue wings towards S2 (position~8). The irradiated side of the bright CO-layer demonstrates the narrowest width of the \co{} and \tco{} lines $\approx 4$\kms{} over the entire map including the area without YSOs deeper to the molecular clump in positions 11-18.

Local star formation reveals itself through the wings of the \co{} line clearly visible in positions 7-11. We find a narrow dip in the \co{} profiles at $-8$~\kms{}, which we associate with a self-absorption effect. The \tco{} line profiles are single-peaked everywhere except for the position~8, where velocities of the \co{} and \tco-dips on the double-peaked profiles are not the same, where the \tco{} dip is shifted to the red by 1~\kms{} and the \tco{} peak coincides with the \co-dip. Therefore, these dips can be caused by different factors: self-absorption for \co{} (also previously detected by \citet{Kirsanova_2019, Figueira_2020, Kabanovic_2022} for the (2--1) and (3--2) transitions, but shifted to the blue side for the (1--0) transitions in \citet{Anderson_2015}) and the gas infall for \tco, while the second statement requires further investigation using tracers of dense gas. 

\subsection{Simulated \co{} and \tco{} line emission}\label{sec:simul}

\begin{figure}
\includegraphics[width=\columnwidth]{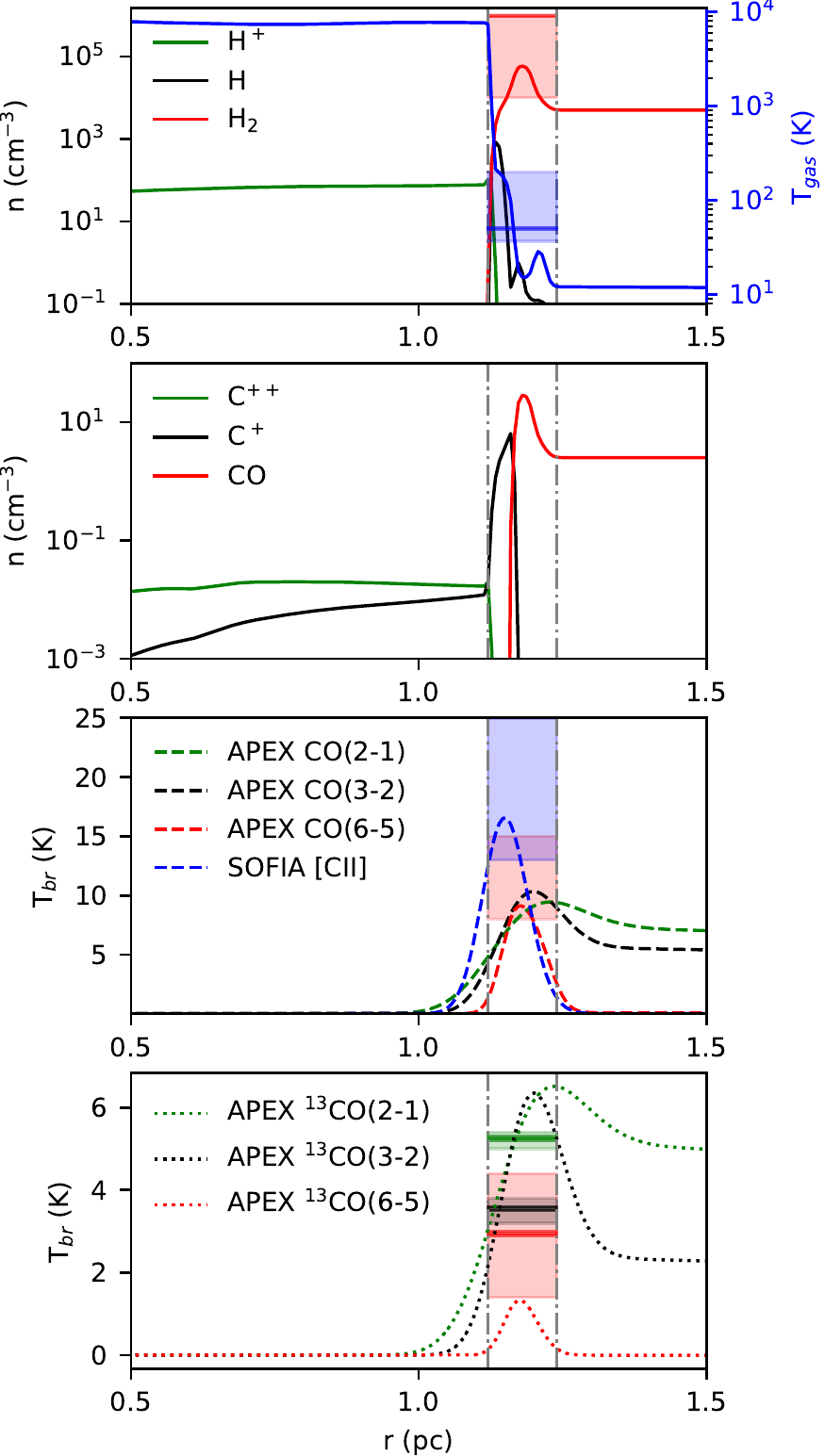}
\caption{Simulated (lines) and observed (colour rectangles and horizontal straight lines) properties of the RCW~120 PDR. Simulated physical conditions and chemical composition in the dense neutral layer around the \hii{} region obtained with the MARION model are shown on the two top panels. The ionizing star is located outside the left border of the panels in the beginning of the horizontal axe. The red rectangles and the red lines show the $\pm \sigma$ significance intervals and the best-fit solution of the non-LTE analysis with the observed and archival data (see Sec.~\ref{sec:analysis}). Simulated brightness of the \cii, CO and $^{13}$CO emission in different lines obtained with the RADEX software is shown in two bottom panels. Green and black horizontal colour bold lines and rectangles show the brightness and $\pm \sigma$ intervals for the archival $^{13}$CO(2--1) and (3--2) lines. The observed range of \co{} and \tco{} brightness are shown by the red rectangle. The \tco{} brightness used for simulations with RADEX is shown by the red horizontal line.}
\label{fig:radexmodel}
\end{figure}

In order to study the origin of the bright CO-layer, we compare the observed \co{} and \tco{} intensities in the layer with results of numerical simulations. Fig.~\ref{fig:radexmodel} shows the edge of the \hii{} region and surrounding PDR, where gas temperature and density change up to several orders. The dense layer of the neutral material, which has been accumulated around the \hii{} region by a shock wave, whose propagation preceded the expanding ionized gas. The shocked region contains the H/H$_2$ and C$^+$/C/CO transition. The mass of the shocked region is mostly contained in the gas where hydrogen and carbon are in the forms of H$_2$ and CO, respectively. Even in the region, where carbon is in the form of C$^+$, $n({\rm H_2})>n({\rm H})$ up to factor of few. Carbon becomes doubly ionized in the \hii{} region and distribution of C$^+$ traces the irradiated border of the PDR. The outer attenuated border is located inside of the shocked layer, there the extinction reaches 10 and higher values. The width of the chocked layer is $\approx 0.1$~pc.

It is seen from Fig.~\ref{fig:radexmodel}, that spatial distribution of brightness temperatures for the CO and $^{13}$CO (2--1) and (3--2) line emission does not allow to resolve the thickness of the shocked layer. The peaks of the (2--1) transitions are shifted outside the PDR, the peak widths are two times broader than the layer, and the contrast between the peak and emission from undisturbed molecular gas is only 20-50\%. The (3--2) transitions better trace the shocked layer, while the widths of the peaks are still broader and the contrast reaches only a factor of 3. Only the (6--5) line emission reaches its maximum exactly at the densest part of the shocked layer and has zero-to-zero width exactly the same as the width of the shocked layer. The peak of the simulated \cii{} emission is shifted to the ionizing star for less than the APEX beamsize at the 690~GHz. Simulated brightness of the \co{} and \tco{} lies in the interval and agrees with the low limit of the observed values, respectively, see Fig.~\ref{fig:obsemission} The simulated \cii{} also lies in the observed interval of values, see contours in Fig.~\ref{fig:obsemission}. Simulated $^{13}$CO(2--1) and (3--2) lines intersect the ranges of the line brightness taken from \citet{Kirsanova_2019}.

Comparing the view of the bright CO-layer in Fig.~\ref{fig:obsemission} and results of the simulations in Fig.~\ref{fig:radexmodel}, we relate this layer with the shocked dense neutral material, which has been shovelled up on the border of the ionized region during hundreds of thousands of years. The shocked layer represents large-scale structure, related with evolution of the \hii{} region. In RCW~120, we see the \co{} emission also outside the shocked layer, probably due to local heating from embedded YSOs. Below we show how the shocked layer is displayed in the kinematic structure  of the observed region.

\section{Physical conditions in molecular gas}\label{sec:analysis}

In this section we analyse radial distribution of excitation conditions for the \co{} and \tco{} lines using the LTE approach and also combine these data with available archival data for the non-LTE analysis in the most interesting directions of the observed region.

\subsection{LTE}\label{sec:lte}

\begin{figure}
    \centering
    \includegraphics[width=\columnwidth]{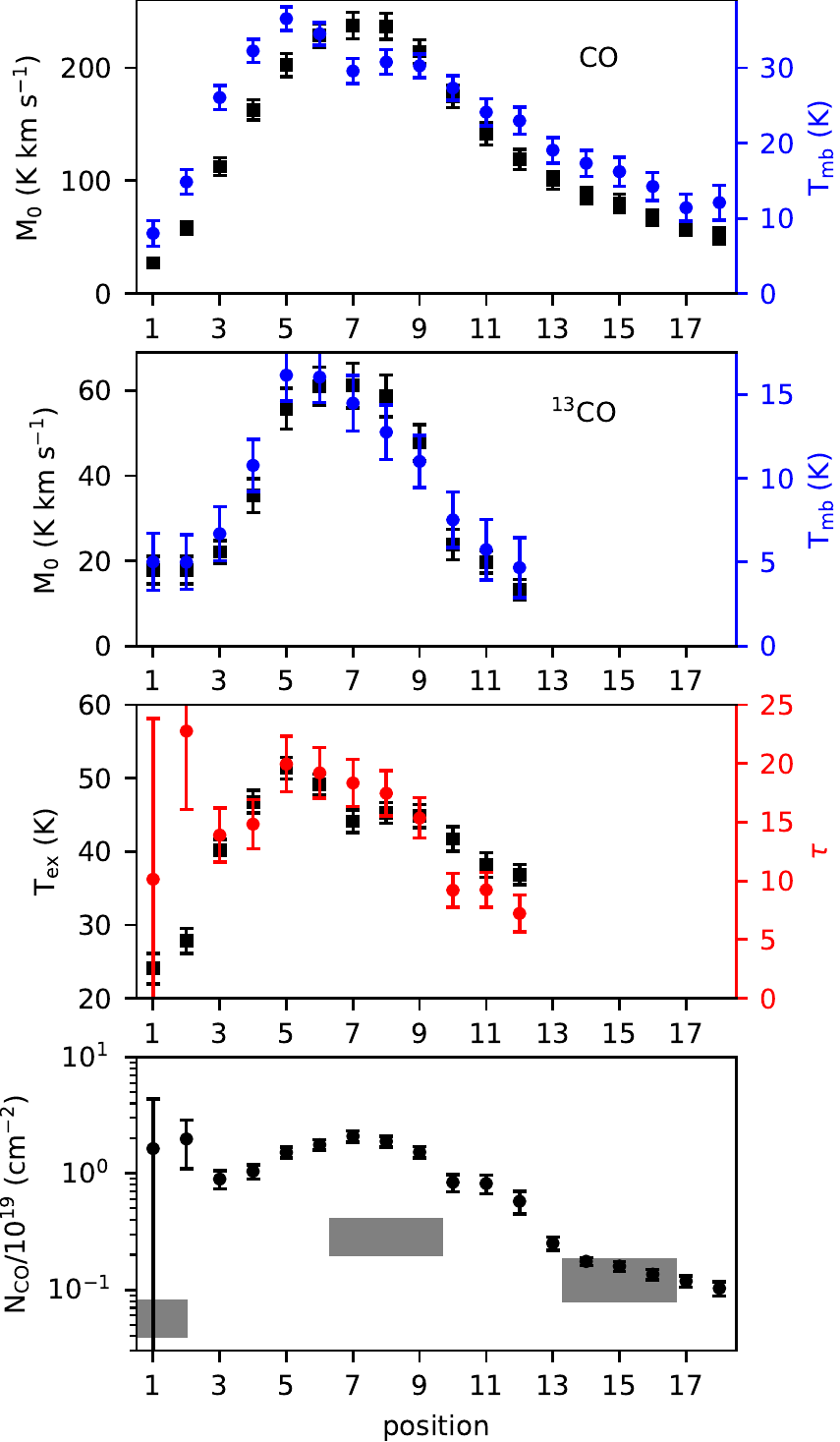}
    \caption{Integrated and peak intensities of the CO(6--5) and $^{13}$CO(6--5) emission (two top panels) as well as the results of the LTE analysis along the cut. Position 8 corresponds to the direction to the Core~2 YSO. Grey rectangles at the bottom panel correspond to the ranges of $N_{\rm CO}$ from the non-LTE analysis.}
    \label{fig:LTEanalysis}
\end{figure}

We show in Fig.~\ref{fig:LTEanalysis} (top panel) radial profiles of the \co{} peak brightness and integrated intensity, taken perpendicular to the PDR along the \tco{} strip. Spatial difference between the maxima of the peak and integrated intensities $\approx9$\arcsec, i.e. about the size of the telescope beam. Spatial behaviour of the \tco{} peak intensities qualitatively resembles the \co{} emission with the sharp/smooth decreasing on the inner/outer sides of the PDR. However, the \tco{} integrated intensity shows almost symmetrical spatial distribution around the YSO~2 which can be related to its lower optical depth. Comparing the observed and simulated radial distributions in Fig.~\ref{fig:radexmodel}, we note that both the \tco{} distributions are symmetric. However, there is no smooth decrease of the simulated \co{} emission on the outer side of the shocked layer. We relate the enhancement of the observed \co{} emission in positions 11-18 with local star-forming activity in the region, which is not included into our simulations.

With both the \co{} and \tco{} lines, along the same strip in molecular cloud, we estimate the optical depth of the \co{} line and the CO column density using an LTE~approach, explicitly described by \citet{Mangum2015},  see Eq.~(79) therein.  This approach assumes the same constant excitation temperature ($T_{\rm ex}$) for \co{} and \tco{} along the line of sight. For the (6--5) transition we used the energy of the upper level $E_{\rm u}=116.2$~K, rotation constant $B_0=55101.014\times10^6$~Hz, and dipole moment $\mu = 0.1$~Debye. Results of the LTE~analysis: optical depth of the \co{} line $\tau$, excitation temperature $T_{\rm ex}$ and CO column density $N_{\rm CO}$ are shown in Fig.~\ref{fig:LTEanalysis}. In order to estimate uncertainties of $T_{\rm ex}$, $\tau$ and $N_{\rm CO}$, we varied integrated intensities of the \co{} and \tco{} lines as well as the \tco{} brightness temperature within their $1\sigma$ intervals. Repeating the procedure 100 times, we find the average values of these parameters with their standard deviations.

The \co{} lines are optically thick in all positions where we detect the \tco{} emission with the \co{} optical depth $\tau \approx 18-20$ in the direction of S2 and $10 \leq \tau \leq 20$ in the direction of the PDR. Dividing the $\tau$ value by the $^{12}$C/$^{13}$C ratio $=61$, suitable for the galactocentric distance of RCW~120~\citep{Wilson_1999}, we find optically thin \tco{} line along the strip even in the direction of the YSO. The value of $T_{\rm ex}$ does not exceed 50~K and has almost flat radial profile along the strip within the interval of $40-50$~K. As we work in the LTE~approximation, we estimate thermal width of the \co{} line using the $T_{\rm ex}$ value and find it $\leq 0.5$~K. Optical depth of the \co{} line also contributes to the broadening by a factor of $\approx {\rm log}(\tau) \approx 2-3$, i.~e. up to 1\kms{}. Totally, thermal and depth-related broadening provide less than 25\% of the observed line widths. Therefore, the widths of both \co{} and \tco{} lines from Fig.~\ref{fig:individual_spectra} are non-thermal, but the width of the self-absorption feature is thermal. Therefore, the self-absorption of these relatively high excitation lines ($E_{\rm u} > 100$~K) can be caused by cool gas between RCW~120 and the observer.

CO column density distribution along the strip has two maximal values of $N_{\rm CO} \approx 2\times 10^{19}$~cm$^{-2}$ in the direction of the PDR and the YSO S2, respectively (spectra 1, 2 and 8 in Fig.~\ref{fig:individual_spectra}).  Using values of hydrogen column density from the ViaLactea survey \citep[based on {\it Herschel} data, see][]{Marsh_2017}, we determine relative abundance of CO in the bright CO-layer and other directions along the strip and find it $\sim 10^{-4}$. The CO/H$_2$ ratio is close to the elemental abundance of carbon, H$_2$ and CO are well mixed with each other in the PDR and in the remaining part of the molecular clump also. Therefore, the CO and H$_2$ dissociation fronts should be situated closer to the ionizing star than position~1 in Fig.~\ref{fig:obsemission}. While the ViaLactea maps may underestimate hydrogen column density, the uncertainty related to the {\it Herschel} fluxes does not significantly affect the densest regions, see, e.~g. a comparison made by \citet{2021MNRAS.506.4447L}.

\subsection{non-LTE}\label{sec:nonlte}

\begin{figure}
    \centering
    \includegraphics[width=0.99\columnwidth]{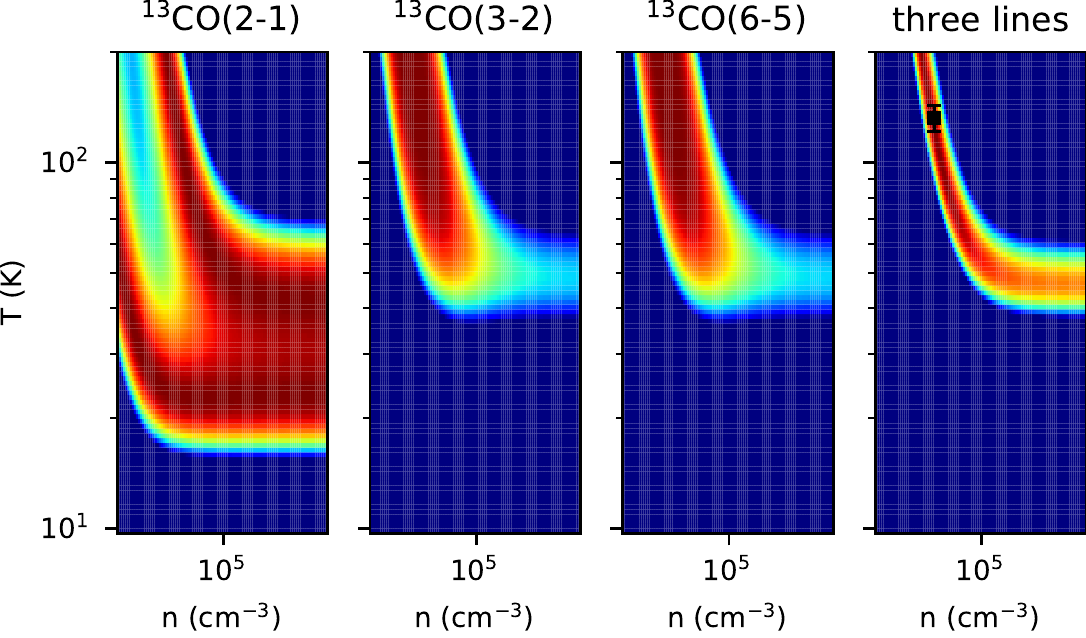}
        \caption{Grid of the $^{13}$CO RADEX models for different gas temperatures and densities at the best-fit $N_{\rm ^{13}CO}=4.7\times 10^{16}$~cm$^{-2}$ for the S2 YSO position. Colour shows $\chi^2$ where values are decreasing towards dark red.}
    \label{fig:nonLTEanalysis}
\end{figure}

While straightforward LTE analysis allows estimating gas temperature ($T_{\rm gas}$) in molecular gas, it doesn't give any information about the gas number density ($n_{\rm H_2}$). Therefore, we combine the \tco{} line emission data from the present study and the $^{13}$CO(2--1) and $^{13}$CO(3--2) data from \citet{Kirsanova_2019}, to estimate physical conditions in three different regions in RCW~120. Namely, towards the S2 YSO, to the PDR layer (position with the offset $\Delta \alpha, \Delta\delta = 20$\arcsec,20\arcsec{}, between positions 1 and 2 in Fig.~\ref{fig:obsemission}), and towards molecular cloud ($\Delta \alpha, \Delta\delta = -20$\arcsec,--20\arcsec{}, between positions 15 and 16), all positions are given relatively to the S2~YSO at $\alpha=17\rm ^h$12$\rm ^m$08.700$\rm ^s$ and $\delta=-38^\circ$30\arcmin46.40\arcsec (J2000). There are no YSOs in the PDR and MC directions, therefore we can obtain parameters of molecular gas without contamination of any activity related to star formation. We fixed the widths of all the CO lines to 4.4\kms{} and background temperature to 2.7~K in order to calculate non-LTE models with RADEX software on a grid of different $T_{\rm gas}$ (from 10 to 200~K divided by 100 intervals), $n_{\rm H_2}$ ($10^3 -10^7$~cm$^{-3}$, 100 intervals), $N_{\rm ^{13}CO}$ ($10^{14}-10^{18}$~cm$^{-2}$, 100 intervals). The best-fit results and $\pm \sigma$ significance intervals, found with the $\chi^2$-test \citep[][]{wall_jenkins_2003} are shown in Table~\ref{tab:nonLTE}. In order to estimate dependence of the $\chi^2$ minimum on the noise level of the various $^{13}$CO lines, we varied their integrated intensities within the $1\sigma$ levels: 0.3, 0.1 and 1.5~K for $^{13}$CO(2--1), $^{13}$CO(3--2) and \tco, respectively. Repeating the procedure by 100 times, we found the average minimum of the $\chi^2$ value and corresponding average parameters with their standard deviations. For the YSO model, the average minimum $T_{\rm gas}$ and $n_{\rm H_2}$ with their standard deviations are shown in Fig.~\ref{fig:nonLTEanalysis}. Comparison of the error bars with the shape of the $\chi^2$ dependency of the $T_{\rm gas}$ and $n_{\rm H_2}$ values shows that the quality of the observational data does not affect on the best-fit parameters, while the degeneracy of the best-fit parameters is significant due to the $^{13}$CO line excitation conditions. However, we become convinced that the solution for the minimum is steady.

We find that the $N_{\rm ^{13}CO}$ value is determined with an accuracy of 30-60\%, while the $\pm \sigma$ significance intervals are much broader for $T_{\rm gas}$ and $n_{\rm H_2}$. Considering only those $T_{\rm gas}$ and $n_{\rm H_2}$, which corresponds to the best-fit $N_{\rm ^{13}CO}$ values, we find almost the same intervals, those are shown in Fig.~\ref{fig:nonLTEanalysis} for the YSO position (the $\chi^2$ diagrams for the PDR and MC positions have similar shape).  The value of $T_{\rm gas}$ from non-LTE analysis agrees within a factor of two with $T_{\rm ex}$ values from the LTE analysis in the PDR and MC positions, but the minimum $\chi^2$ value was found at $T_{\rm gas} = 124$~K towards the YSO. While the values of $N_{\rm CO}$ from LTE and non-LTE analysis are also in agreement within the uncertainty intervals (see the bottom panel in Fig.~\ref{fig:LTEanalysis}, where we show $N_{\rm CO} = 61\times N_{\rm ^{13}CO}$), the non-LTE value is lower by an order of magnitude towards S2. Therefore, the non-LTE effects of the line excitation are important towards the S2 YSO, but the LTE approximation is valid for the physical conditions in the MC position.  The gas temperature $>100$\,K, found in S2, agrees with the results by \citet{Kirsanova_2021}, who found hot gas towards the YSO.

The non-LTE intervals for the $n_{\rm H_2}$ and $T_{\rm gas}$ values at the PDR position agrees also with the simulated values in the shocked material, see Fig.~\ref{fig:radexmodel}. This agreement confirms that the bright CO-layer is the shocked gas indeed. The best-fit $n_{\rm H_2}$ value in the PDR position is by 2-2.5 orders of magnitude higher than that in the YSO and MC positions. We explain this fact by compactness of dense material in the direction of the YSO and relatively large beam size of our observations. The minimum non-thermal line width of the \co{} line together with the high density of the PDR also indicate that the bright CO-layer might represent the gas where internal chaotic motions were suppressed by the shock.

\begin{table}
\centering
\begin{tabular}{c|c|c|c}
\hline
           & $T_{\rm gas}$     & $n_{\rm H_2}$ & $N_{\rm ^{13}CO}$    \\
           &  (K)              & ($10^4$~cm$^{-3}$)   &($10^{16}$~cm$^{-2}$)        \\
\hline
PDR        &  $50_{37}^{200}$  & $950_{1.0}^{1000}$ &  $0.8_{0.6}^{1.1}$ \\
\vspace{0.025cm}\\
Core~2 YSO & $124_{40}^{200}$ & $1.3_{0.6}^{1000}$ &  $4.7_{4.3}^{6.1}$\\ 
\vspace{0.025cm}\\
MC         &  $64_{22}^{200}$ & $1.5_{0.2}^{1000}$ &  $1.7_{1.3}^{2.8}$  \\
\hline
\end{tabular}
\caption{The best-fit results and $\pm \sigma$ significance intervals of non-LTE analysis for the Core~2 YSO, PDR and molecular cloud.}
\label{tab:nonLTE}
\end{table}

\section{Kinematics of the molecular gas}\label{sec:kinematics}

\begin{figure*}
    \centering
    \includegraphics[width=0.85\columnwidth]{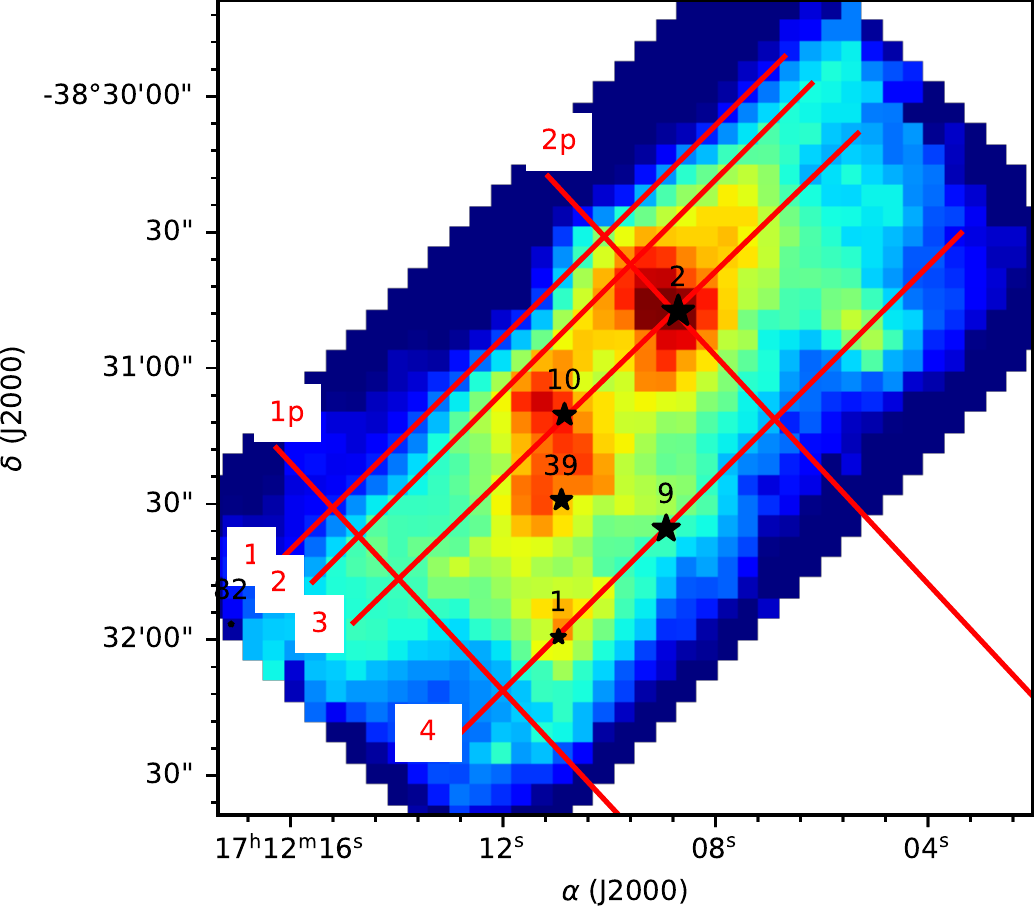}
    \includegraphics[width=0.59\columnwidth]{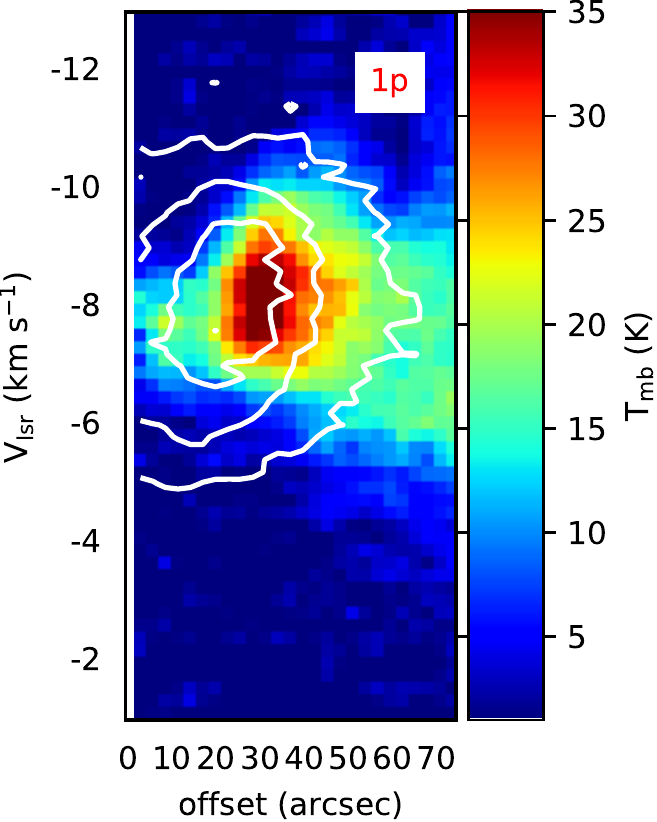}
    \includegraphics[width=0.59\columnwidth]{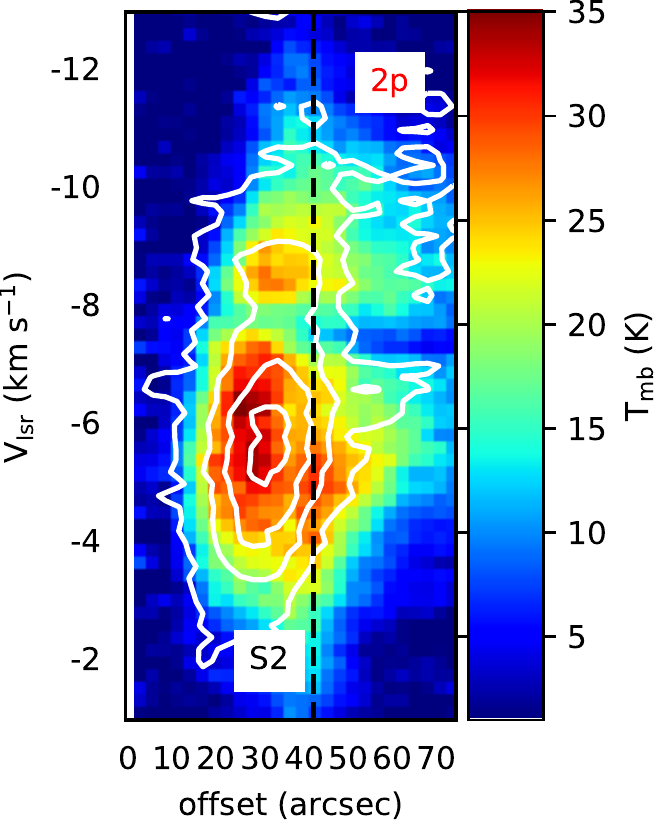}\\
    \includegraphics[width=\columnwidth]{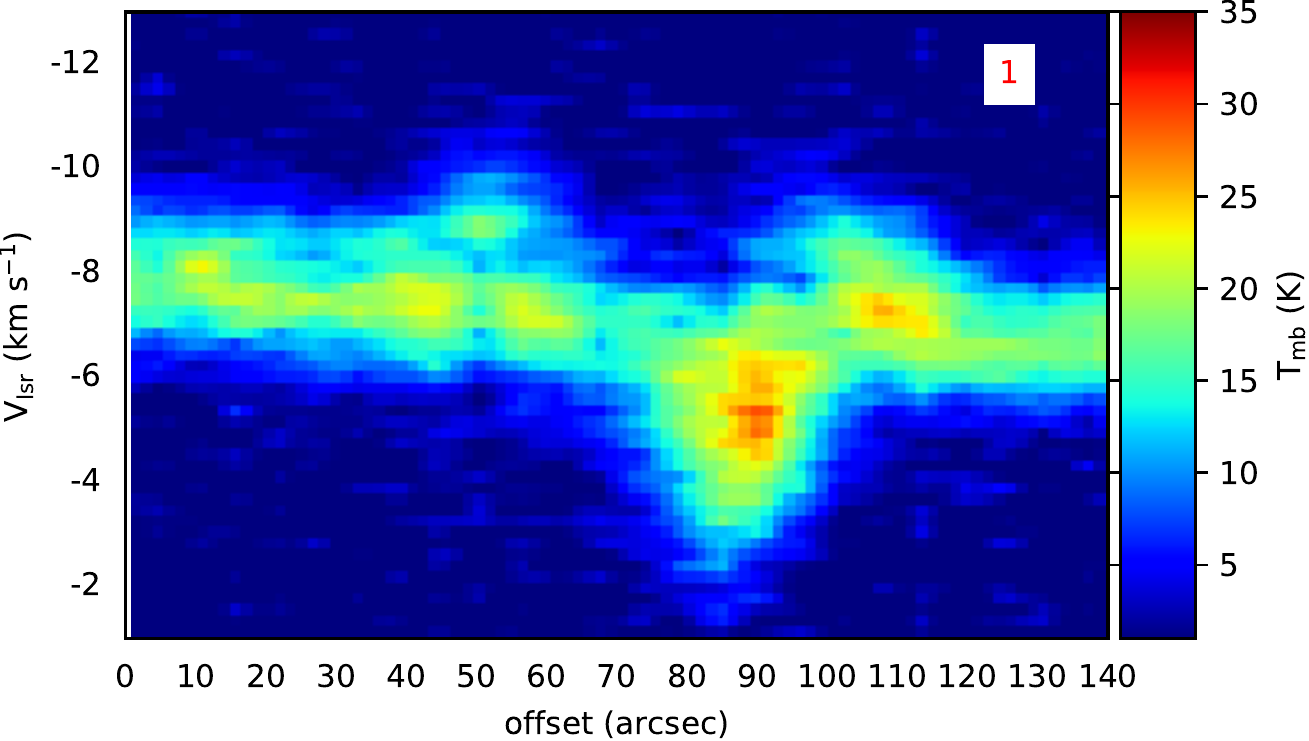}
    \includegraphics[width=\columnwidth]{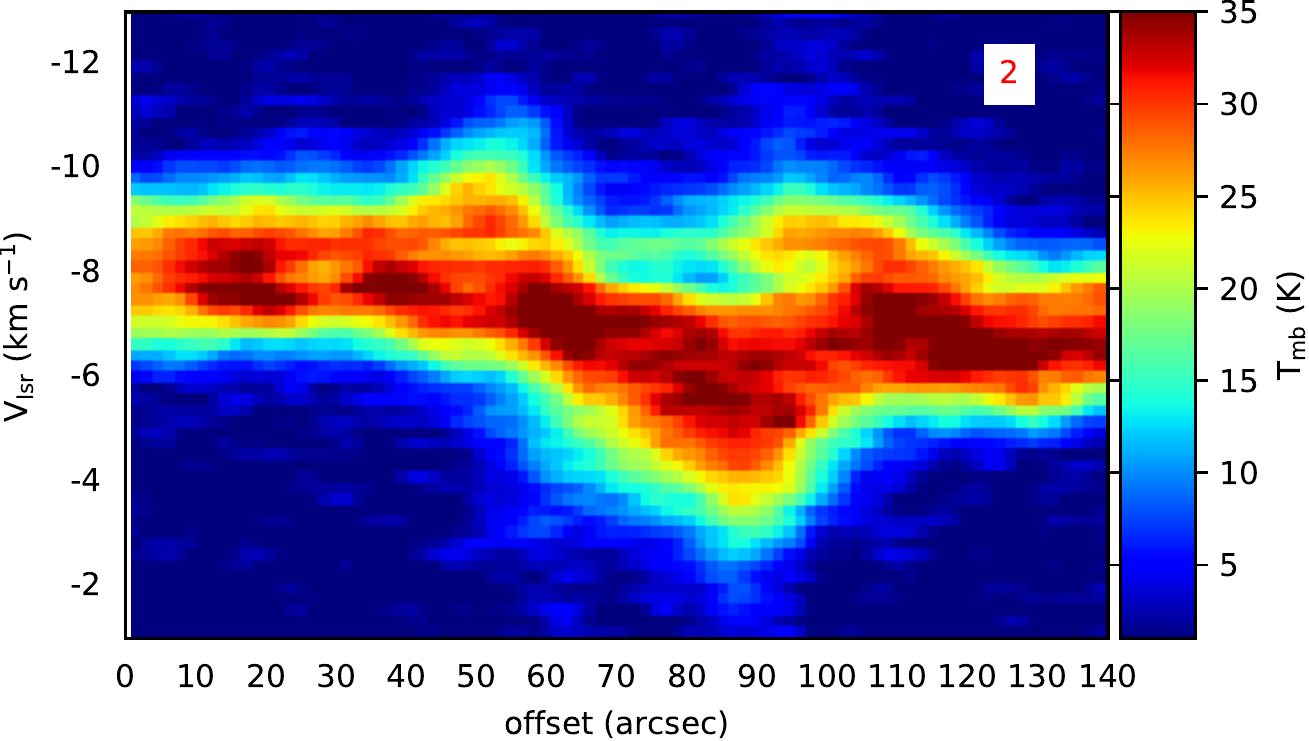}\\
    \includegraphics[width=\columnwidth]{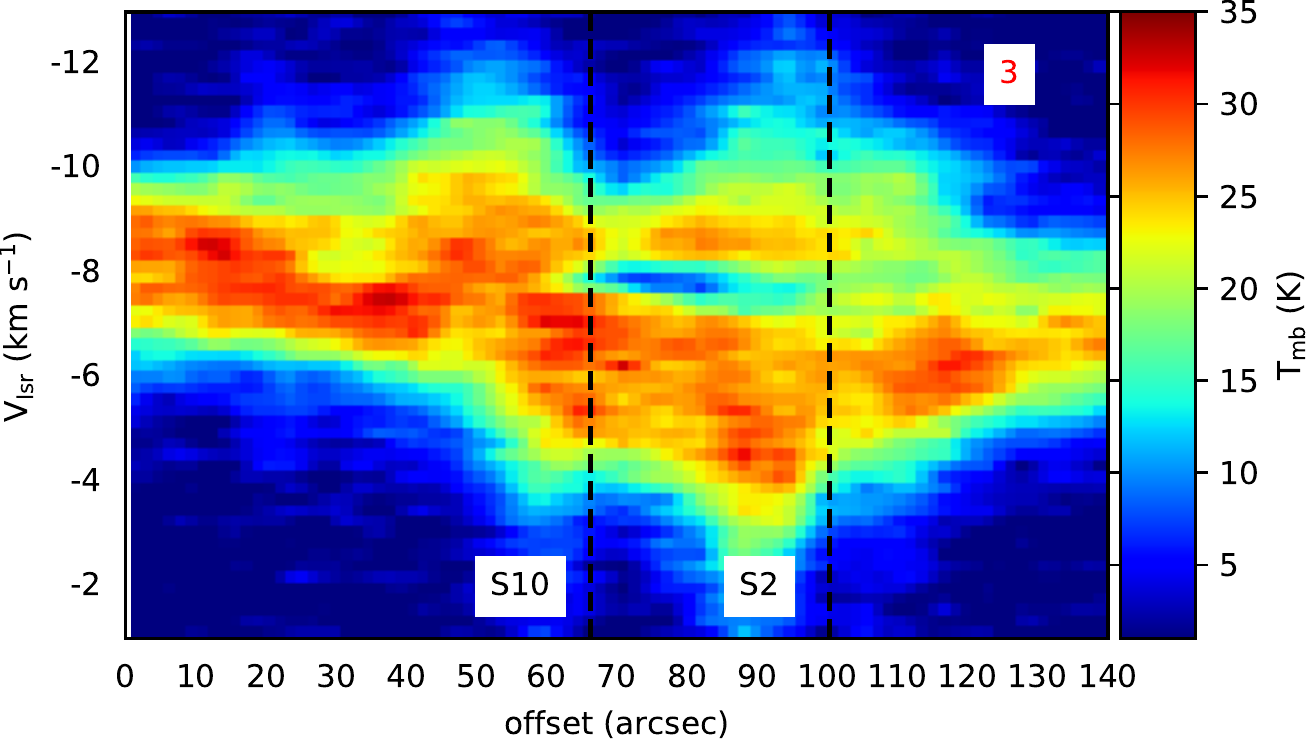}
    \includegraphics[width=\columnwidth]{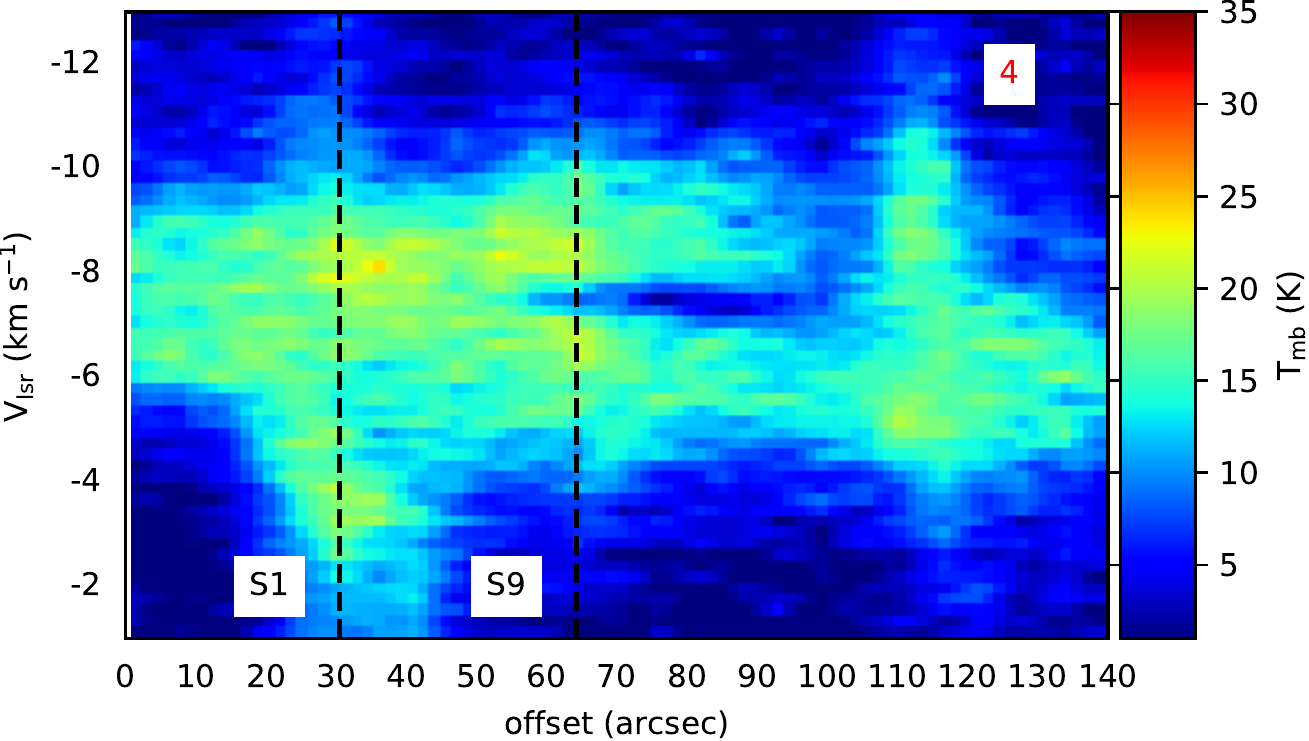}
    \caption{Pv~diagrams of the CO(6--5) line emission. Top row: map of the integrated \co{} emission where red lines show directions of the pv~diagrams. Pv~diagrams for the \co{} (color) and [\cii] lines (white contours), taken perpendicular to the ionization front are given on two right panels (1p and 2p). Middle and bottom rows: pv~diagrams taken parallel to the ionization front (1--4). Black dashed lines show positions of YSOs on the pv~diagrams, names of those are given in black.}
    \label{fig:pv_perpendicular}
\end{figure*}

Position-velocity (pv) diagrams in Fig.~\ref{fig:pv_perpendicular}, made parallel and perpendicular to the ionization front, show complex structure of the observed region. Making the cuts perpendicular to the ionization front, we clearly see the bright CO-layer with narrow \co{} lines, which we relate to the shocked layer. This gas is unperturbed by the embedded YSOs, see e.~g. the diagram~1p at the offset~$\approx 30$\arcsec. The lines become broader at the larger offsets $\approx 50-70$\arcsec{} as we move away from the PDR to the molecular clump. All the observed line widths are non-thermal (see analysis of physical parameters in Sec.~\ref{sec:analysis}), therefore we suggest that the velocity dispersion can be decreased in the shocked layer due to compression (and higher turbulence decay) and star-forming activity leads to the broadening of the lines outside the layer. The pv~diagram~2p demonstrate local gas kinematics related to star-forming activity in the S2~YSO, namely broad lines with wings at the offset~$\approx 30-50$\arcsec{} and less broad at the higher offsets deeper into molecular clump. We note a narrow dark line at $V_{\rm lsr} =-7.5$\kms{}, which we relate to self-absorption effect caused by the large-scale foreground material extended over RCW~120 and described in Sec.~\ref{sec:intro}. Stratified structure of PDR and a transition from atomic to molecular gas can be seen from comparison of the \co{} and archival [\cii] pv~diagrams, also shown in Fig.~\ref{fig:pv_perpendicular}. We find the [\cii] peak closer to the ionization front than the \co{} peak at the pv~diagram~1p. However, the [\cii] and \co{} peaks coincide at diagram~2p. We will discuss properties of the C$^+$/CO transition of the PDR in Sec.~\ref{sec:discussion}.

Pv~diagrams, made parallel to the ionization front, distinguish regions of the large-scale shocked layer and dense molecular clump. The diagrams~1 and 2 show a straight line at --8\kms{} which makes a red-shifted arch up to --2\kms{} at the offsets~60--110\arcsec. The arch becomes double-sided in diagrams~3 and 4 between the same offsets. The broadest \co{} lines appear in positions of YSOs. Comparing the diagrams~3 and 4 with the line profiles in Fig.~\ref{fig:individual_spectra}, we relate these broad lines to outflows to the red and blue sides, accompanying the star formation process towards the YSOs. Therefore, from these pv~diagrams we distinguish two regions: the first one is regular structure related to the large-scale shocked layer from the expanding RCW~120 (at the offsets $<60$\arcsec{} and $>110$\arcsec) and the second one related to star-forming regions in the dense clump.

\begin{figure}
    \centering
    \includegraphics[width=\columnwidth]{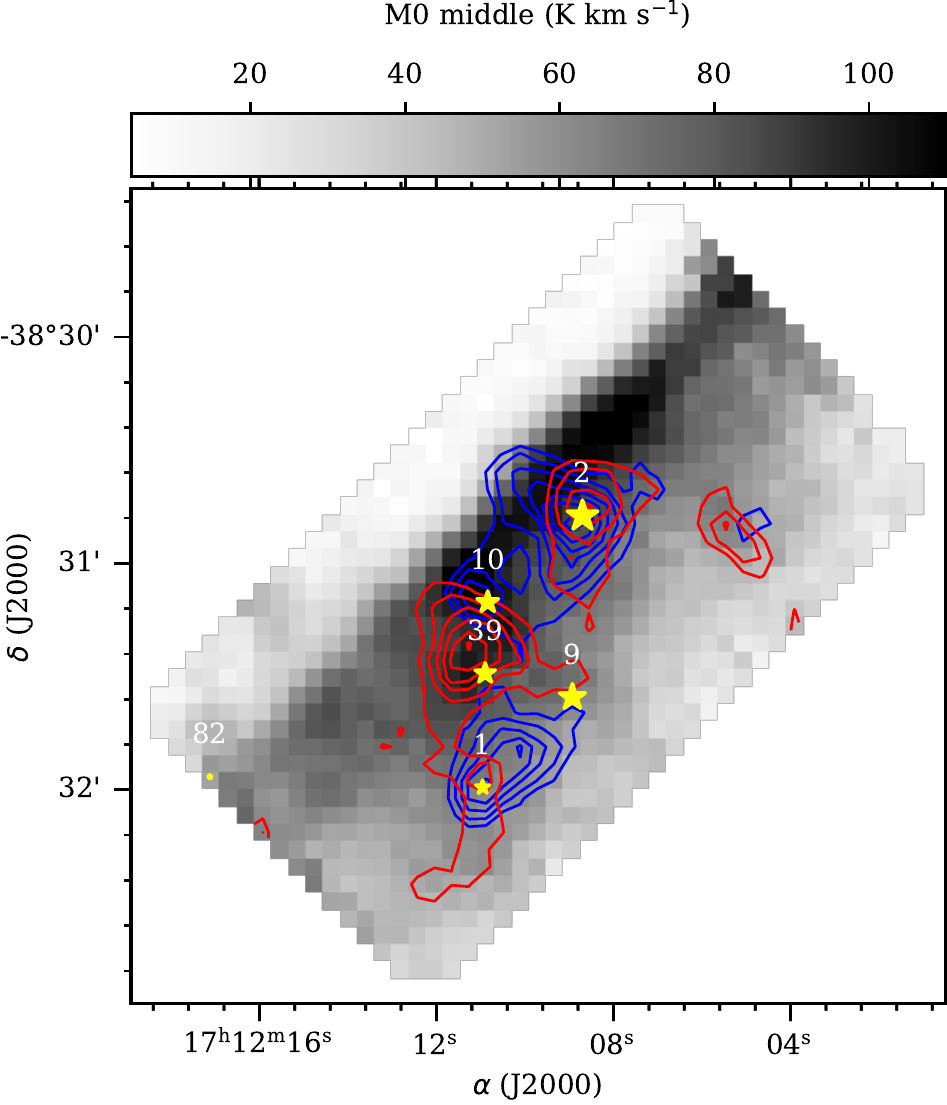}
        \caption{Channel maps of the \co{} emission around RCW~120. There are three velocity intervals in this map: $V_{\rm lsr} < -9$ (blue contours), $-9 < V_{\rm lsr} < -4$ (middle, shown by greyscale) and $V_{\rm lsr} > -4$~\kms{} (red contours). The contours are shown from 30 to 90~K~\kms{} through 10~K~\kms.}
    \label{fig:outflowsreal}
\end{figure}

Outflow candidates, detected by us in the present study, do not look like those that were found previously by \citet{Figueira_2020} with low excitation CO(3--2) line. Therefore, we show the candidates in Fig.~\ref{fig:outflowsreal} considering the same velocity intervals as it was done for the CO(3--2) line. We find not one but several candidates using two times higher angular resolution. The first candidate, related with the S2~YSO, appears as an outflow oriented along the line of sight as the red and blue lobes almost coincide. Different geometry visible with the higher excitation lines can be related with the complex structure of the outflow near and far from the exciting source, traced by the (6--5) and (3--2) transitions, respectively. We find intersecting red and blue patches toward the S1~YSO where the blue-shifted part is much brighter than the red one. The red wings of methanol lines were seen previously by \citet{Kirsanova_2021}, therefore we do not have doubts about the outflow activity there. We deduced from emission maps of CH$_3$OH, SiO and DCO$^+$ molecules (Plakitina et al., in prep.), that only red and blue patches around S1 and S2 are real outflows, but the other ones visible between S9, S10 and S39 are parts of complex kinematic structure of the region.

\section{Discussion}\label{sec:discussion}

We know that stellar wind \citep{Luisi_2021} and ionized gas \citep[][]{Sanchez-Cruces_2018} expel all neutral material and produce dense expanding PDR around the bubble. The appearance of a dense layer, swiped by the shock wave, in the neutral side of PDR accompanies the expansion \citep[][see also calculations by e.g.~\citet{Hosokawa_2005, Kirsanova_2009, Akimkin_2015,  2018arXiv180101547B}]{Spitzer_1978}. We associate our bright CO-layer with the shocked molecular layer, moving away from the ionizing star, as we have shown above. While the bubble and the PDR resembles a sphere, the molecular region certainly has different geometry and resembles a flattened envelope oriented face-on which can be modelled in the plane-parallel approach. 

Abundance of CO in the shocked layer agrees with the elemental value for this part of the Galaxy, it means that carbon is locked in this gas-phase CO. Since the position of the H$_2$ dissociation front normally coincides with the bright 8\micron{} emission from PAHs \citep[e.~g.][]{2010ApJ...725..159F, 2021ApJ...919...27K}, we conclude that the CO and H$_2$ dissociation fronts are situated on the inner border of the bright \co{} layer. The merged fronts appear as a natural consequence of the ongoing propagation of the dissociation fronts through the dense molecular cloud \citep[see e.~g. simulations by][]{Storzer_1998, Hosokawa_2005, Kirsanova_2019b}: the denser the gas the closer the fronts. The pv~diagrams from Fig.~\ref{fig:pv_perpendicular}, taken perpendicular to the ionization front, confirm the merging of the H$_2$ and CO dissociation fronts in the dense material. For instance, comparing the diagram~1p and 2p, taken outside and through the dense material, respectively, we find different separation between the [\cii] and \co{} peaks. The peaks do not coincide in the diagram~1p, but they merge in the diagram~2p. Taking into account divergence of the \co{}, [\cii] and 8~\micron{} peaks in the south-eastern part of the observed region, we suggest that the CO and H$_2$ dissociation fronts might be separated there.

The shocked molecular gas appears as the most regular structure in our maps of the line emission in Fig.~\ref{fig:obsemission} and pv~diagrams in Fig.~\ref{fig:pv_perpendicular}. While the line width is still non-thermal, its value is about a factor of 2 lower in the shocked layer than in the molecular clump. Significant broadening of the molecular lines occurs near YSOs and is related to outflows. The shock does not disperse the dense clump unlike the outflows. This scenario looks similar to the recent findings by \citet{Kavak_2022} in the Orion region and can be widespread around O-type stars surrounded by younger star-forming regions. Comparing the positions of YSOs found by \citet{Deharveng_2009, Figueira_2017} with the location of the dense shocked layer (see Fig.~\ref{fig:spitzer}) we agree with conclusion by \citet{Figueira_2020} that the YSOs were pre-existed before the shock wave from RCW~120 reached the molecular clump.

\section{Conclusions}\label{sec:conclustion}

In this study, we made spatially and velocity-resolved observations of the RCW~120 PDR and nearby molecular gas with \co{} and \tco{} lines and found a bright CO-emitting layer towards the PDR. We simulated physical conditions and emission of several CO and C+ lines in the RCW~120 PDR and found reasonable agreement with the observed values. Based on this agreement, we related the CO-bright layer with shocked molecular gas moving away from the ionizing star due to expansion of \hii{} region. While the shock itself is not visible, we found its impact on the neutral material around RCW~120 in our \co{} map. We distinguished between large-scale structures and kinematical features such as the shocked gas and local ones such as dense environments of embedded protostars and outflows. We found the shocked layer as the most regular structure on the \co{} map and in the velocity space, whereas the gas around YSOs was dispersed by the outflows.

\section*{Acknowledgements}

We are thankful to S.~V.~Salii and K.~V.~Plakitina for fruitful discussions. We also thank unknown referee for his/her very relevant comments.

Based on observations with the Atacama Pathfinder EXperiment (APEX) telescope. APEX is a collaboration between the Max Planck Institute for Radio Astronomy, the European Southern Observatory, and the Onsala Space Observatory. Swedish observations on APEX are supported through Swedish Research Council grant No 2017-00648.

M.~S.~K. and Ya.~N.~P. were supported by RSCF, project number 21-12-00373. D.~A.~S. acknowledges financial support by the Deutsche Forschungsgemeinschaft through SPP 1833: "Building a Habitable Earth" (SE 1962/6-1). A.~F.~P. acknowledges the support of the Russian Ministry of Science and Education via the State Assignment Contract no. FEUZ-2020-0038.

\section*{Data Availability}

Calibrated data on \co{} and \tco{} emission generated in the study can be found in the ESO Archive (projects O-0103.F-9301A-2020 and O-0107.F-9318A-2021). Results of the calculations with MARION are available in Zenodo, at https://zenodo.org/record/7447109.



\bibliographystyle{mnras}
\bibliography{CO65_paper}

\bsp	
\label{lastpage}
\end{document}